\documentclass{article}
\usepackage[utf8]{inputenc}

\usepackage{jheppub}

\usepackage{graphicx}
\usepackage[titletoc]{appendix}
\usepackage{xspace}
\usepackage{ulem}

\usepackage[style=numeric-comp,sorting=none]{biblatex}
\title{The HEP Software Foundation Community}
\author{The HEP Software Foundation\\
\vspace*{4mm}
Contact editors:\\
\vspace*{2mm}
Graeme A Stewart, CERN (\href{mailto:graeme.andrew.stewart@cern.ch}{graeme.andrew.stewart@cern.ch})\\
Peter Elmer, Princeton University (\href{mailto:Peter.Elmer@cern.ch}{Peter.Elmer@cern.ch})\\
Elizabeth Sexton-Kennedy, Fermilab (\href{mailto:sexton@fnal.gov}{sexton@fnal.gov})
}
\date{April 2022}

\begin{document}

\noindent
\begin{tabular*}{\linewidth}{lc@{\extracolsep{\fill}}r@{\extracolsep{0pt}}}
 & & HSF-DOC-2022-01 \\
 & & May 17, 2022 \\ 
 & & Copyright (C) 2022 CERN, Princeton and Fermilab, licence CC-BY-4.0 \\
 & & \\
\end{tabular*}
\vspace{2.0cm}

\abstract{The HEP Software Foundation was founded in 2014 to tackle common problems of software development and sustainability for high-energy physics. In this paper we outline the motivation for the founding of the organisation and give a brief history of its development. We describe how the organisation functions today and what challenges remain to be faced in the future.}

\maketitle

\hypertarget{history}{%
\section{History}\label{history}}

Over the past 50 years, the experimental particle, nuclear and
astroparticle physics communities have iteratively evolved a significant
amount of community structure. This was a natural result of the growing
size, scale and time duration of the experiments and the centralisation
of facilities at large laboratories. National, and now international,
collaborations are typically required to build, operate and maintain the
large detectors used in these experiments. No single university or
laboratory can provide all of the necessary expertise and required
effort. The largest collaborations have grown from 100s of collaborators
in the 1990s to 1000s at (for example) the Large Hadron Collider (LHC)
at CERN. This community has also developed a broad ecosystem of
methodologies and technologies that are used to build successive
experiments and upgrades to existing experiments. While a specific
instrument can necessarily only be used for a single experiment at any
given time, the large commonalities in methodology should permit the
development of research software which can be used widely in the
community for different projects. Despite this, much of the software
development remained somewhat siloed within individual experiments or,
at most, one or another host laboratory, with only a few exceptions.

This is not to say that in the history of HEP there was no common software. 
CERNLIB~\cite{CERNLIB} was a foundation library written in the Fortran era and used by many experiments.  Elements of it have been rewritten in C++ and constitute some of the most widely used software packages in the field.  Projects such as ROOT, Geant4, and various generators have effectively acted as common glue for many experiments.  However in the software layers above these foundation and toolkit libraries, redundant
solutions that are difficult to evolve and sustain over time
(years or decades for large experiments!) are common. 
To a large extent, software speed, performance and efficiency had been ignored previously, because the costs due to inefficient software could be ignored in the past. Software is as much an intellectual product as well as a tool, thus a new approach was needed.

First steps in the direction of collaborating on modernising, in 2011-2012, led to the formation of a
cross-experiment ``Concurrency Forum''~\cite{CERN-RD-MULTICORE} to discuss
the specific software challenges brought by changes in microprocessor
technology (multi-core, wide vectors, GPUs). Driven initially by CERN
and Fermilab, the forum demonstrated community interest in wider
software collaborations. By 2014-2015, a number of colleagues involved
in HEP software for many years were discussing a more ambitious and
broader scope for research software collaborations in HEP. This
eventually led to the formation of the \textbf{High-Energy Physics (HEP)
Software Foundation (HSF)}. The driving motivations for this initiative
were that the physics upgrades anticipated in the coming decades,
particularly the High-Luminosity LHC~\cite{HL-LHC}, would put enormous
pressure on the software used in HEP; that much of our software was
already decades old; the changes in microprocessor technology brought
new challenges to the table; and that there was an urgent need to train
new talent and to attract investment to the field, which could be better
supported when common, multi-experiment, efforts were promoted. More
generally, additional community structure which promotes research
software collaborations, not tied to single experiments or laboratories,
has greater potential to enhance the longer term sustainability of the
software.

The very first workshop~\cite{CERN-WS} attempted to build on the
community experience within the large experiments, however there was too
much discussion of ``governance'' questions. Individual experiments need
to operate a large well-integrated detector, manage pooled resources
(such as computing and storage) and at the end of the day produce
scientific publications signed by the entire collaboration. Thus
governance questions within experiments are critical. This top-down
approach was less obvious for the envisioned research software
collaborations, which can be more ``ecosystem-like''. It also made
engaging experiments of very different sizes more challenging. By the
end of the workshop most participants had concluded that a different
structure was needed. Subsequent workshops, one in North America and
one in Europe to aid inclusivity~\cite{SLAC-WS,LAL-WS}, brought together
many HEP experiments, HEP specific software projects and other non-HEP
organisations, such as the Apache Software Foundation and the Software
Sustainability Institute. Here, the principle of a ``do-ocracy'' was
enshrined, to encourage activity from the bottom-up, with developers
leading, which was a far more productive approach to community building.

From these early workshops the idea was born of preparing a Community
White Paper (CWP), laying out the roadmap for software and computing in
HEP in the 2020s. As a way to fortify these community-led efforts, the
Worldwide LHC Computing Grid (WLCG) gave a formal charge to the HSF to
prepare this paper~\cite{CWP-Charge}. Many individuals volunteered and
the US National Science Foundation was an early investor with dedicated
funding to help organise and carry out CWP workshops~\cite{SDSC-CWP,LAPP-CWP}. These workshops were focal points of an intense year of
identifying key domain areas and potential solutions for the field and
setting up working groups who would prepare topic-specific white papers.
This approach was able to greatly broaden the involvement of key
individuals and solidified the active communities around the HSF. Once
the working groups had produced their domain-specific white papers, an
editorial team took charge of synthesising a unified single version of
the white paper, which was published with extremely wide community
support, 310 signing authors from 124 institutes~\cite{Albrecht2019}.

The CWP was not the only activity happening in the HSF at this time.
Particularly where there was an identified gap between different
communities, be these inter-experiment, between projects, or even
between theory and experiment, the HSF was ideally placed to bridge gaps
and bring different people together. Workshops on analysis ecosystems~\cite{ANALYSIS-ECO} 
and on computational aspects of event generators~\cite{COMP-GEN} were notable examples. In addition the HSF was a natural
forum for considering more radical software developments and their
potential, gathering experiment and expert feedback on progress and
possibilities~\cite{GEANTV-RD}.

\hypertarget{hsf-activities}{%
\section{HSF Activities}\label{hsf-activities}}

While much of the early HSF activity was focused on software community
building through the CWP, working groups were started where common
topics were obviously identified within the community, e.g., in the
domain of packaging and distributing HEP software stacks. These groups
brought together experts and interested parties and focused around
technical discussions.

In the wake of the CWP it was clear that this model would work very well
for the areas of most concern for the future, so the model was broadened
and, over a few years, eight working groups were established with about
3 conveners in each case, appointed annually and with a nomination
system that allows both stakeholder (experiment, institution) input and
bottom-up volunteers for running these groups. In order to marshall
these activities it was necessary to have some critical amount of
binding effort, so the decision of CERN management to allow significant
(0.5 FTE) time from one person to the HSF was crucial and has had a key
multiplication effect.

Where these groups are involved in areas that are `traditional', e.g.,
detector simulation or reconstruction, there is a strong involvement
with ongoing developments in the experiments and in well established
software projects. The focus is the exchange of ideas and discussions of
common problems. In a number of cases the HSF had identified topics of
interest that were simply not covered elsewhere in the field and then
the HSF working group has had a further leading and catalysing effect.
This is particularly the case for the use of Python in HEP, led by the
PyHEP working group; and for computational aspects of physics event
generators, led by the Generators working group~\cite{Valassi2021}.

In addition to being a focus for the exchange of ideas and techniques,
when WGs identify a concrete topic where a paper can usefully be
prepared, the HSF is a natural place for organising pan-experiment input
and encouraging some standardisation (e.g., in analysis level metadata
or in detector conditions data access). This has been recognised by more
formal bodies outside the HSF, such as WLCG and the LHCC, who often ask
the HSF to marshall community inputs for updates on development
activities and reviews of development plans. This is an important point
in terms of recognising the contribution of the organisation. In turn,
this fosters recognition for the contribution of our individual members
and helps their careers to advance.

This engagement with strategic bodies in the field, where the HSF
advocates for software investment~\cite{stewart_graeme_a_2018_2413005}, leads the HSF to
work in tandem with other funded software R\&D projects in HEP, where
the HSF will support funding applications and then work with these
projects, enhancing their connection to the community and the impact of
their work. Practically projects can contribute through the working
groups closest to their R\&D areas.

The HSF also engages with other like minded bodies and collaborates on
regular series of meetings regarding accelerator programming or software
and computing for nuclear physics and also organises itself as an
umbrella organisation, with CERN, to run Google Summer of Code for HEP
software projects and experiments.

In the past the HSF organised face-to-face workshops and the intention
is to restart such activities once the pandemic passes, but in the
meantime virtual workshops have played a role. In some circumstances
these can even have a greater impact, with the PyHEP workshops in 2020
and 2021 registering more than 1000 participants~\cite{PYHEP20,PYHEP21}.
In large part this reflects a strong didactic element in the Python
area. This thread is reflected also in that the HSF, together with
IRIS-HEP, SIDIS, The Carpentries and the ROOT project~\cite{IRIS-HEP,SIDIS,CARPENTRIES,Brun1996,ROOT}, has put a strong emphasis on training
activities~\cite{Malik2021} and now runs regular training events in
fundamental software skills and in C++ programming. This is seen as a
critical activity in the field and attempts are now also being made to
have these activities as feeders to encourage trainees to be involved in
software development and training.

\hypertarget{outcomes-and-conclusions}{%
\section{Outcomes and
Conclusions}\label{outcomes-and-conclusions}}

Contrary to initial ideas, the HSF has not, by and large, run software
projects themselves - without actually having resources to disburse it
is better to allow such projects to be independent and work with the HSF
as is useful. That said, HSF events have proved to be fertile ground for
people from different backgrounds to meet (e.g., nuclear and
astroparticle physics) and even to start common software projects that
then take on a life of their own. Fostering funded projects has led to
new investment in software R\&D in HEP and a higher recognition of the
importance of promoting excellent software developers in their careers;
this is a major success that was a direct outcome of the CWP process.

The HSF has now established itself as a recognised part of the HEP
software landscape where it links strategic bodies to the community of
software developers. It remains a challenge to continue to build the
next generation of HEP software developers and make them feel involved
and part of an organisation like the HSF, but work to improve training
and engage younger colleagues through the working group process is hoped
to improve this. Looking forward to post-pandemic activities, where
face-to-face interactions can happen again, will also help to continue
to build HEP software communities.


\sloppy
\raggedright
\clearpage
\printbibliography[title={References},heading=bibintoc]

\end{document}